\documentclass[prb,aps,twocolumn,floats,showpacs,amsmath,amssymb]{revtex4}
\usepackage{bm,graphicx}
\usepackage{color}
\newcommand{\be}{\begin{equation}}
\newcommand{\ee}{\end{equation}}
\newcommand{\bea}{\begin{eqnarray}}
\newcommand{\eea}{\end{eqnarray}}

\begin{document}

\title{Dimensionality effects on the Edwards polaron: From one to three dimensions}
\author{M. Chakraborty$^1$}
\author{N. Mohanta$^2$}
\author{A. Taraphder$^{1,3}$}
\author{B.I. Min$^4$}
\author{H. Fehske$^5$}
\affiliation{$^1$Department of Physics, Indian Institute of Technology, Kharagpur, India}
\affiliation{$^2$Center for Electronic Correlations and Magnetism, Theoretical Physics III, 
Institute of Physics, University of Augsburg, 86135 Augsburg, Germany}
\affiliation{$^3$Center for Theoretical Studies, Indian Institute of Technology, Kharagpur, India}
\affiliation{$^4$Department of Physics, 
Pohang University of Science and Technology, Pohang, 790-784, Korea}
\affiliation{$^5$Institut f\"{u}r Physik, Ernst-Moritz-Arndt-Universit\"{a}t 
Greifswald, 17487 Greifswald, Germany}

\date{\today}


\begin{abstract}
Employing a self-consistent (optimized) variational diagonalization scheme we investigate the formation of polaronic quasiparticles in a spinless fermion-boson transport model that couples the movement of charge carriers to fluctuations and correlations of a background medium. The background is parameterized by bosonic degrees of freedom. The variational fermion-boson Hilbert space is constructed to achieve high accuracy in one to three spatial dimensions  with modest computational requirements. To characterize the ground-state properties of the Edwards model in the single-particle sector,  we present exact numerical results for the polaron band dispersion, quasiparticle weight, Drude weight, mass enhancement, and the particle-boson correlations in a wide parameter regime. In the Edwards model, transport will be quasi-free, diffusive or boson-assisted in the weakly fermion-boson coupled, fluctuation-dominated or strongly correlated regimes, respectively. Thereby correlated transport is not only assisted  but also limited by the bosonic excitations. As a result, the Drude weight remains finite even in the limit of very small boson frequencies. For a strongly correlated background, closed loops are important, in any dimension, to generate a finite effective particle mass even when the free fermion has an infinite mass. 
\end{abstract}

\maketitle
\section{Introduction}
The question of how a background medium affects the motion of a charge carrier is one of the most heavily debated issues in solid state physics. In this connection the background may typify a great variety of situations. It could, for example, represent a simple deformable lattice~\cite{La33,Pe46a},  
or a highly correlated Mott insulator~\cite{LNW06,Be09,WOH09}. In the former case, the mutual interaction between the charge carrier and the lattice deformation may constitute a new quasiparticle, the so-called (lattice) polaron~\cite{Fi75}, an electron dressed by a phonon cloud. Here, depending on the nature of the particle-phonon coupling~\cite{SS93}, non-polar short-ranged or polar long-ranged, (small) Holstein~\cite{Ho59a,Ho59b} or (large) Fr\"ohlich~\cite{Fr54,Fr74} polarons will form, with distinct transport and optical properties~\cite{Alex07,IRF06,FT07,AD10}. In the latter case, the undoped (insulating) parent compounds develop magnetic, orbital, or charge ordered phases at low temperatures~\cite{TNFS00}. 
Prominent examples are the three-dimensional (3D) ferromagnetic (colossal magnetoresistive) manganites~\cite{JTMFRC94}, the quasi-2D antiferromagnetic (high-temperature superconducting) cuprates~\cite{BM86}, the 2D transition metal dichacogenides (competition between unconventional superconductivity and charge-density-wave order)~\cite{KLVT14}, or the 1D halogen-bridged (charge-density-wave) transition-metal complexes~\cite{BS93}. 
Doping such systems, the charge carriers, electrons or holes, cannot propagate freely as their motion normally disturbs 
the spin-, orbital-, or charge-order of the background. Nevertheless coherent particle transfer may occur on a 
reduced energy scale: this time the particles have to carry a cloud of background (e.g., spin or orbital) excitations. 
The corresponding quasiparticles are frequently called in the literature spin~\cite{KLR89,MH91a} or orbital polarons~\cite{WOH09}.

The Edwards fermion-boson model constitutes a paradigmatic model to describe quantum transport and polaron formation for such situations~\cite{Ed06}.  
It is based only on a few, very plausible assumptions: (i) as a charge carrier moves along a transport path in a solid,  
it creates an excitation with a certain energy in the background medium at the site it leaves or annihilates an existing excitation at the site it enters, (ii) because of quantum fluctuations, excitations in the 
background may appear and disappear spontaneously, and 
(iii) the (de)excitations of the background can be parameterized as bosonic degrees of freedom. 
In this way, the model captures,  to varying degrees, some of the basic aspects of the 
Holstein-, $t$-$J$-, Hubbard- or Falicov-Kimball-model physics. The Edwards Hamiltonian reads 
\begin{equation}
 H=H_{b} -\lambda\sum_i(b_i^{\dagger}+b_i^{})
              +\omega_0\sum_i b_i^{\dagger}b_i^{}\,,
\label{model1}
\end{equation} 
where the first term, $H_{b}$=$-t_{b}\sum_{\langle i, j \rangle} f_j^{\dagger}f_{i}^{}
 (b_i^{\dagger}+b_j^{})$, describes a boson-affected nearest-neighbor hopping $(\propto t_b)$ of spinless fermionic particles $(f_i^{(\dagger)})$, the second term allows for the relaxation $(\propto \lambda)$ of the bosons ($b_{i}^{(\dagger)}$), and the third term gives the energy $(\propto \omega_0)$ of the bosonic background excitations. We note that in the Edwards model, the coupling between fermions and bosons notably differs from that in the Su-Schrieffer-Heeger (SSH) model~\cite{SSH79,CSG97,MDCBNPMS10} where the modulation of the electronic hopping is given by the difference of the on-site lattice displacements $(\hat{X_i}-\hat{X}_j)$, with $\hat{X_i}\propto (b_i^\dagger +    b_i^{})$ and  $b_i^\dagger$ creating a phonon at site $i$. Self-evidently, the Edwards fermion-boson coupling also differs from the local Holstein
electron-phonon interaction~\cite{Ho59a,Ho59b}, which is to a quantized (dispersionless) optical normal mode of lattice vibration. In the Edwards model, the (Einstein) boson simply accounts for the (de)excitation of the background, through which the fermion is moving. 

So far the Edwards model could only be solved in 1D, namely by numerical approaches like exact-diagonalization and density matrix renormalization group (DMRG) techniques. There, for a single particle, quasi-free, diffusive, or correlated transport emerges~\cite{AEF07}. The latter sets in at small $\lambda$ and large $\omega_0$ when the background becomes ``stiff'', a case that resembles the motion of a hole in an antiferromagnetic spin background~\cite{Tr88,KLR89,MH91a,EEAF10}.
At half-filling, $\tfrac{1}{N} \sum_{i}\langle f_i^{\dagger}f_{i}^{}\rangle$=$1/2$,  a metal-insulator quantum-phase transition 
has been proven to exist: Entering the strongly correlated regime a repulsive Tomonaga-Luttinger liquid gives way to a charge-density-wave ground state~\cite{WFAE08,EHF09}. Off half-filling, attractive Tomonaga-Luttinger-liquid phases and regions with phase separation have been detected~\cite{ESBF12}. In 2D, the treatment of the Edwards model was only approximate. 
In the single-particle sector, using the momentum-average approach~\cite{BF10}, the quasiparticle dispersion throughout the Brillouin zone has been calculated. Very recently, employing the projective renormalization method~\cite{CBFBS16}, a tendency towards unconventional superconductivity has been observed for the 2D half-filled band case.

In this paper we focus on Edwards polaron formation in the single-particle sector. Since the microscopic structure of the Edwards polaron is rather diverse, with---depending on the model parameters---lattice polaron or spin polaron characteristics, we utilize a self-consistent variational numerical diagonalization technique~\cite{CCM07,CDC11,CMCD12,CM13,CTM14} to address this issue in one to three spatial dimensions. Due to the huge bosonic Hilbert space, the dimensionality effects on the Edwards polaron problem have not been studied before. The proposed method is capable of computing the band dispersion, the quasiparticle weight, the effective mass, the 
Drude weight and the spatial particle-boson correlations of the polaron in 1D to 3D. Thereby we particularly investigate how the new energy scale of ``coherent'' particle transport develops. 

That the Edwards model actually captures {\it two} transport channels, a free-fermion hopping channel on a reduced energy scale and the original boson-affected one, becomes already visible performing an unitary transformation, $b_i\to b_i +\lambda/\omega_0$, which eliminates the boson relaxation term. Omitting the constant energy shift $N\lambda^2/\omega_0$ ($N$ is the number of lattice sites), we obtain
\begin{equation}
 H\to H=H_{f} +H_{b}
              +\omega_0\sum_i b_i^{\dagger}b_i^{}\,,
\label{model1}
\end{equation} 
where $H_{f}$=$-t_f\sum_{\langle i, j \rangle} f_j^{\dagger}f_{i}^{}$ with $t_f$=$2\lambda t_b/\omega_0$.
The physics of the Edwards model is governed by two parameter ratios: $t_f/t_b$ (relative strength of free and boson-affected hopping) and $(\omega_0/t_b)^{-1}$ (rate of bosonic fluctuations). In this way $H$ perfectly describes the interplay of ``coherent'' and ``incoherent'' transport channels realized in many condensed matter systems. In what follows we measure all energies in units of $t_b$.

The paper is organized as follows. Section II briefly introduces our numerical approach. In Sec. III,  
we determine the ground-state and spectral properties of the Edwards model and discuss 
several issues of the Edwards polaron problem, especially the dimensionality effect. 
Section IV gives a brief summary and contains our conclusions.

\section{Numerical Approach}
\begin{table}[b]
\begin{tabular}{|c|c|c|cc|c|}
     \hline
     \hline
 $t_f$ & $\omega_0$ & $k$ & $E_0$(SC-VED) &[basis size]& $E_0$(VED) \\  \hline
  20 & 0.5 & $0$ & -40.5922 &(2000000) & -40.591  \\ \hline
  20 & 0.5 & $\pi$ & -40.05 &(2000000) & -40.01  \\ \hline
  2 & 0.5 & $0$ & -5.427354  &(1250000) & -5.42734  \\ \hline
  2 & 0.5 & $\pi$ & -5.020042 &(1500000) & -5.02  \\ \hline
  5 & 2.0 & $0$ & -10.388823488 & (1250000) & -10.388823488  \\ \hline
  5 & 2.0 & $\pi$ & -8.386998 &(1250000)  & -8.38  \\ \hline
  1 & 2.0 & $0$ & -2.59317697703908 & (750000) &  -2.59317697704   \\ \hline
  1 & 2.0 & $\pi$ & -0.8637159668  &(1500000) & -0.86371596   \\ \hline
      \hline
\end{tabular}
\caption {Ground-state energy in a certain $k$ sector for the single-particle Edwards model in 1D. SC-VED results are compared with data obtained 
by the VED approach (which is basically the same as used in Ref.~\onlinecite{AEF07}). Within VED, a variational basis of $18054141$ states is used.
The numerical accuracy is specified in such a way, that the ground-state energies of the $(N_h-1)$-th shell and the $N_h$-th shell match up to the digit presented. For the SC-VED this means that these digits do not change in going from the penultimate to the final iteration cycle.
Given the dimension of the basis and the computational effort, the accessible precision of the data strongly depends on the model parameters   and the momentum.}
\label{t1}
\end{table}

A variational basis is constructed by diagonalizing the Edwards fermion-boson model numerically, starting with a state of a bare electron and adding new states by repeated application of the Hamiltonian, say $N_h$ times. All translations of these states on the infinite lattice are included. Hereafter, we refer to such variational approaches based on exact diagonalization as VED~\cite{BTB99,BT01,CCM07,CDC11,CMCD12,CM13,CTM14,PCT16}. We will also apply 
a self-consistent VED (SC-VED) scheme, which has successfully been used to investigate the 
(extended) Holstein model~\cite{CM13,CTM14}. In the SC-VED framework, we first generate a relatively 
small basis set and calculate the ground-state energy and the wave function.
Then the states with highest probability were identified and the basis is optimized 
by only applying the Hamiltonian on the chosen (highly probable) states. Accordingly, the size of the basis is increased. 
Then the ground-state energy and the wave function are calculated again. This process is continued self-consistently by increasing the basis size at each cycle till the desired accuracy in the ground-state energy is obtained. 
To test the accuracy and efficiency of the SC-VED method, we have recalculated the ground-state energy for a single electron in the 1D Edwards model, a problem that has been solved previously~\cite{AEF07}. Table~I demonstrates the high precision of the SC-VED data, in spite of using a much smaller basis space. For comparison, the VED results included in Table~I were obtained within a variational space of $18054141$ states, corresponding to 
$N_h=18$. Actually, the SC-VED scheme gives even a lower ground-state energy. Note that keeping constant the computational effort, the numerical accuracy of our data depends on the model parameters, as well as on the momentum. Similarly to the Holstein model~\cite{BTB99}, a higher precision  can be reached with less resources if the number of phonons involved is smaller. For the Edwards (Holstein) model this is the strong-correlation (weak-coupling) case, realized at small $\lambda$ and  large $\omega_0$  (small polaron-binding-energy phonon-frequency ratio). That one achieves a lesser accuracy for large momenta was observed for Holstein polaron model as well~\cite{BTB99}. The reason is  the extent of the lattice deformation (size of the polaron), which increases as $k$ approaches the Brillouin-zone boundary, thereby making  any finite-cluster calculation 
more susceptible to finite-size effects. 

\begin{figure}[b]
\includegraphics[scale=0.4]{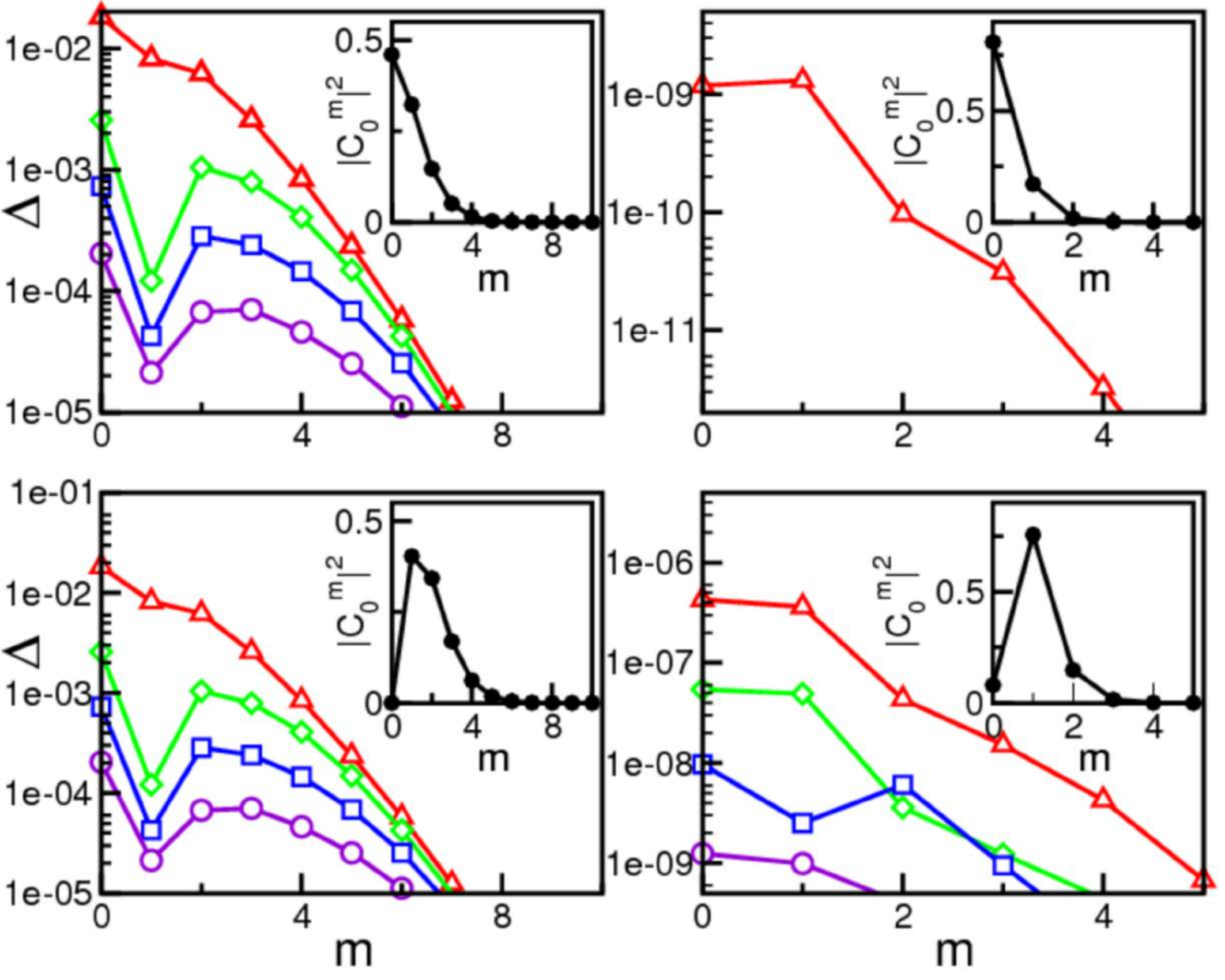} 
\caption{ (Color online) Main panels:  Pointwise convergence  of the weight of the $m$-boson state in the ground state, $|C_{0}^{m}|^{2}$,   
for $k$=$0$ (top) and $k$=$\pi$ (bottom), where $t_f=20$, $\omega_{0}=0.5$ (left) and $t_f=1$, $\omega_0=2$ (right). $\Delta$  specifies the absolute
value of the difference between the first (red triangles up), second (green diamonds), third (blue squares), fourth (violet circles) iteration step 
and the final result  obtained---with the requested accuracy--after five iterations. Insets: Converged values of $|C_{0}^{m}|^{2}$ (filled black circles).}
 \label{f1}
\end{figure}

To ensure that the basis contains an adequate number of bosons for a given parameter set, the weight of the $m$-boson state in the ground state, $|C_{0}^{m}|^{2}$ (for definition, see Refs.~\onlinecite{BWF98,FLW00}), has to be  monitored. The main panel of Fig.~\ref{f1} illustrates the convergence of $|C_{0}^{m}|^{2}$ in the course of the VED iteration process. Beyond that, one recognizes that most bosons are required in the limit of small boson frequencies $\omega_0$ (see inset). We note by now that the limit of $\omega_0\to 0$ substantially differs from the adiabatic limit of the Holstein polaron model~\cite{AFT10}, in that the fermions in the Edwards model do not couple to an (optical) phonon which leads to a static lattice displacement as the frequency of the vibrational mode goes to zero.

\begin{figure}[b]
\vskip -0.0cm
\includegraphics[scale=0.33]{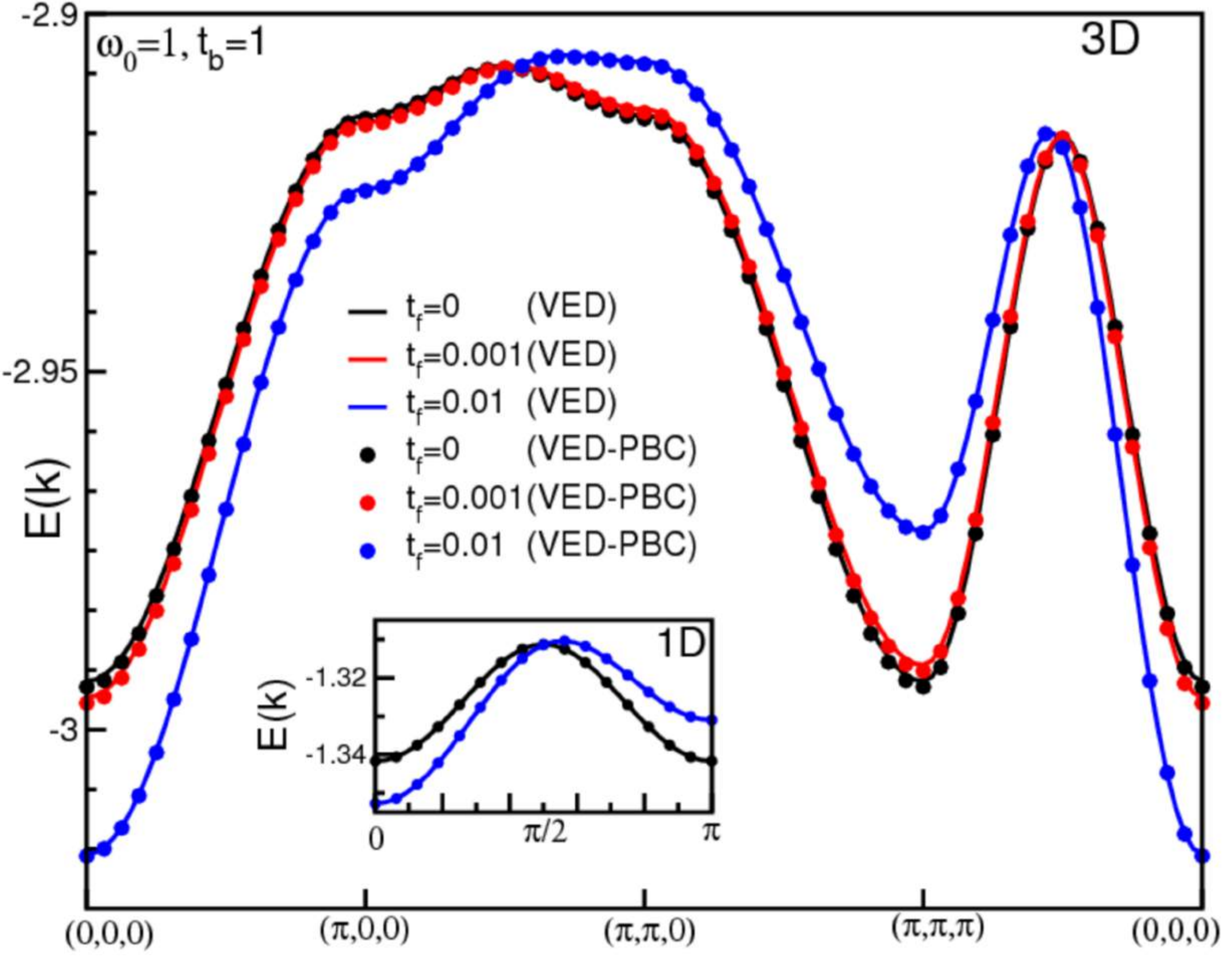}
\caption{(Color online)
Comparison of polaron dispersion with VED and VED-PBC basis set for 3D and 1D (inset) Edwards polaron for $t_b=1$ and $\omega_0=1$ for
small values of $t_f$. $N_h= 8$ basis set has been used for 3D polaron with basis sizes $1755748$ for VED and $1500868$ for
VED-PBC (on a $9\times 9$ lattice). The basis size for 1D Edwards with VED is $18054141$ ($N_h$=$18$) and for VED-PBC is $41485$ ($N_h$=$11$).
The basis size for $N_h$=$11$ with VED is $41488$.  }
 \label{fcom}
\end{figure}

Let us emphasize that  the VED method  allows for a (de facto) continuous variation of the momentum $k$. This is because 
 all translations of the basis states, generated by ``acting'' $N_h$ times with the ``off-diagonal'' hopping and fermion-boson coupling terms 
of the Hamiltonian on an initial root state, on an infinite lattice are included~\cite{BTB99,FT07}.   Treating the 1D Edwards model with $N_h=18$
means that a single bosonic excitation 18 lattice sites away from the fermion is still taken into  account. 
That is why a small Edwards (Holstein or SSH) polaron never feels the boundary in reality.  This advantage of the VED  persists 
in the SC-VED scheme~\cite{CM13}. What happens if we now apply periodic boundary conditions (PBC)? 
Generating the VED basis set on a 1D lattice with $35$ sites and PBC, the Edwards polaron, will be unaffected by 
the boundary conditions until $N_h$=$17$. The PBC comes first into play at the $18^{th}$ basis generation  step, but even here, states having 
18 bosons but no bosonic excitation at the boundary remain unaffected.  This argument holds also in higher dimensions, albeit to a weaker extent. 
Constructing an $N_h=9$  basis set on a 2D $9\times 9$ lattice  with open and periodic boundary conditions, a difference arises 
at $N_h=5$. To substantiate our reasonings, Fig.~\ref{fcom} shows the Edwards polaron band dispersion for the 3D 
and 1D (inset) cases, comparing  the VED and VED-PBC schemes. Apparently, the data match very well:  In 3D (1D) the first 3 (9) decimals agree. 
Sincethe physically most important processes take place in the immediate vicinity of the polaronic quasiparticle, the smaller the radius of the Edwards polaron the better is the agreement of the approaches. Hence the VED-PBC  based on small lattices becomes highly efficient,  whenever the Edwards polaron is rather small, i.e., the background medium is strongly correlated.

\begin{figure}[h]
\vskip -0.0cm
\includegraphics[scale=0.4]{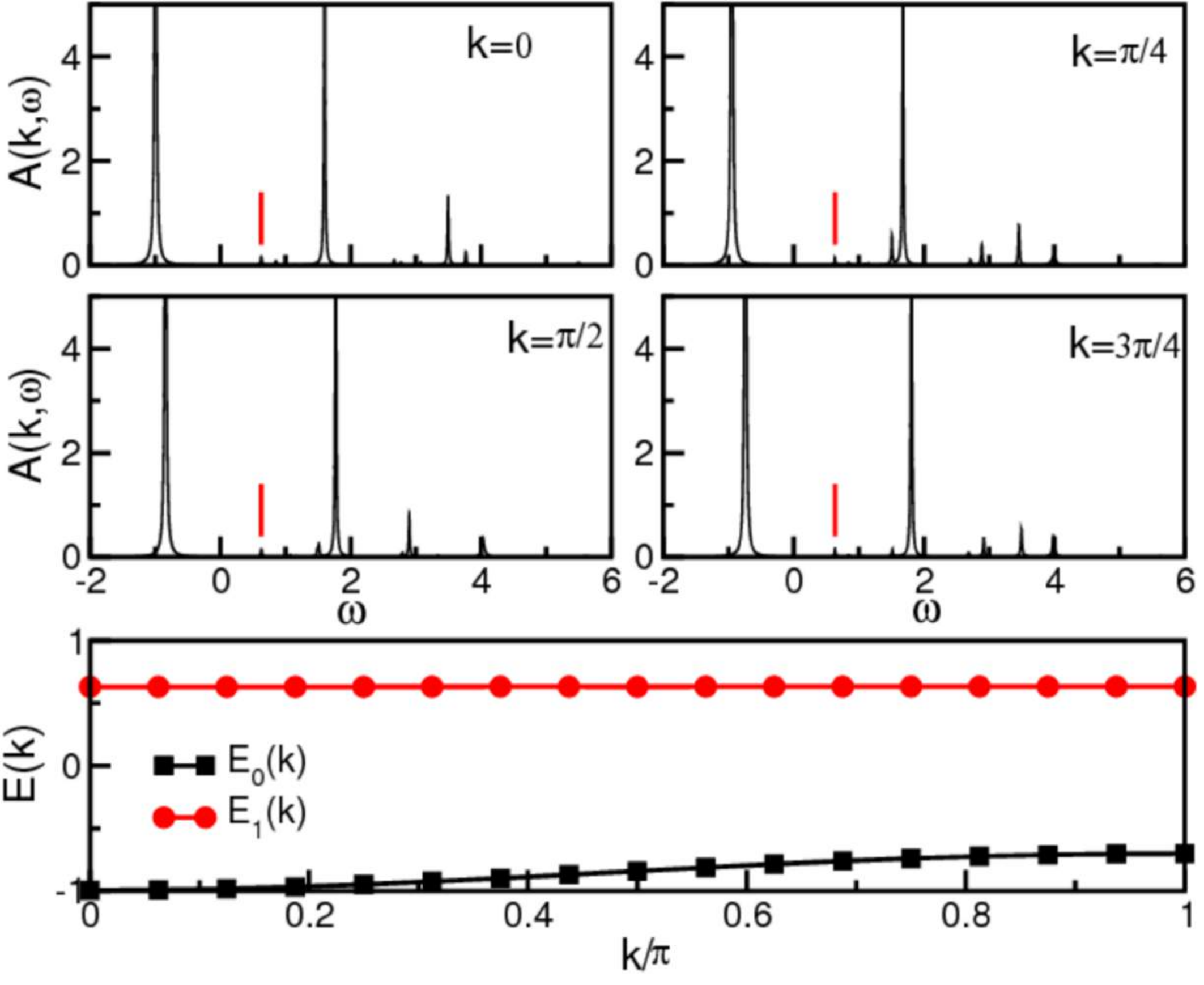} 
\caption{(Color online)
Spectral function $A(k,\omega)$ (top) and band dispersion (bottom) of the 1D Edwards model in the first Brillouin zone. Results are given for $t_f=0.1$ and $\omega_0=2$. In the upper panels, vertical red lines indicate the position of the first excited state (located at $\omega\simeq0.63$). All calculations were performed with a VED basis set of  $18054141$ states ($N_h=18$).}
 \label{f2}
\end{figure}

Next, just to show that our numerical scheme also admits the calculation of excited states and spectral properties, Fig.~\ref{f2} displays the dispersion $E_{1}(k)$ of the first excited state [besides those of the ground state $E_{0}(k)$], and the single-particle spectral function, 
\begin{equation}
A(k,\omega)=\sum_n |\langle \psi_n^{(1)} | f_k^\dagger |0\rangle |^2 \delta(\omega-\omega_0)\,,
\label{ako}
\end{equation}
in the strongly correlated regime. In Eq.~\eqref{ako}, $|\psi_n^{(1)}\rangle $ is the wave function of the $n$-th excited state in the one-particle sector and $|0\rangle$ is the vacuum. Since the particle motion in this parameter regime is essentially determined by the boson-assisted hopping, we find
well separated peaks in the spectrum of all the selected $k$-sectors ~\cite{AEF07}. Of course, 
the ground-state band dispersion follows those of the first peak in $A(k,\omega)$. 
Note that the peak corresponding to the first excited state has only tiny spectral weight, and therefore 
is hardly visible in the spectral function.

Concerning the computational resources, our 1D VED single-electron calculation  takes a basis constructed with $N_h$=$18$. Then, for a lattice with $37$ sites, the matrix dimension is of the order of $18$ millions. For comparison: In 2D (3D), we will take $N_h$=$10$ (8), which means a dimension of about $11$ millions for a $9\times 9$ ($5\times 5 \times 5$) lattice.
In what follows we employ the SC-VED scheme to obtain a better convergence 
(in the $k$=$0$ sector) for all spatial dimensions.

\section{Results and Discussion}
\subsection{Polaron band dispersion}
\begin{figure}[b]
\vskip 0.5cm
\includegraphics[scale=0.4]{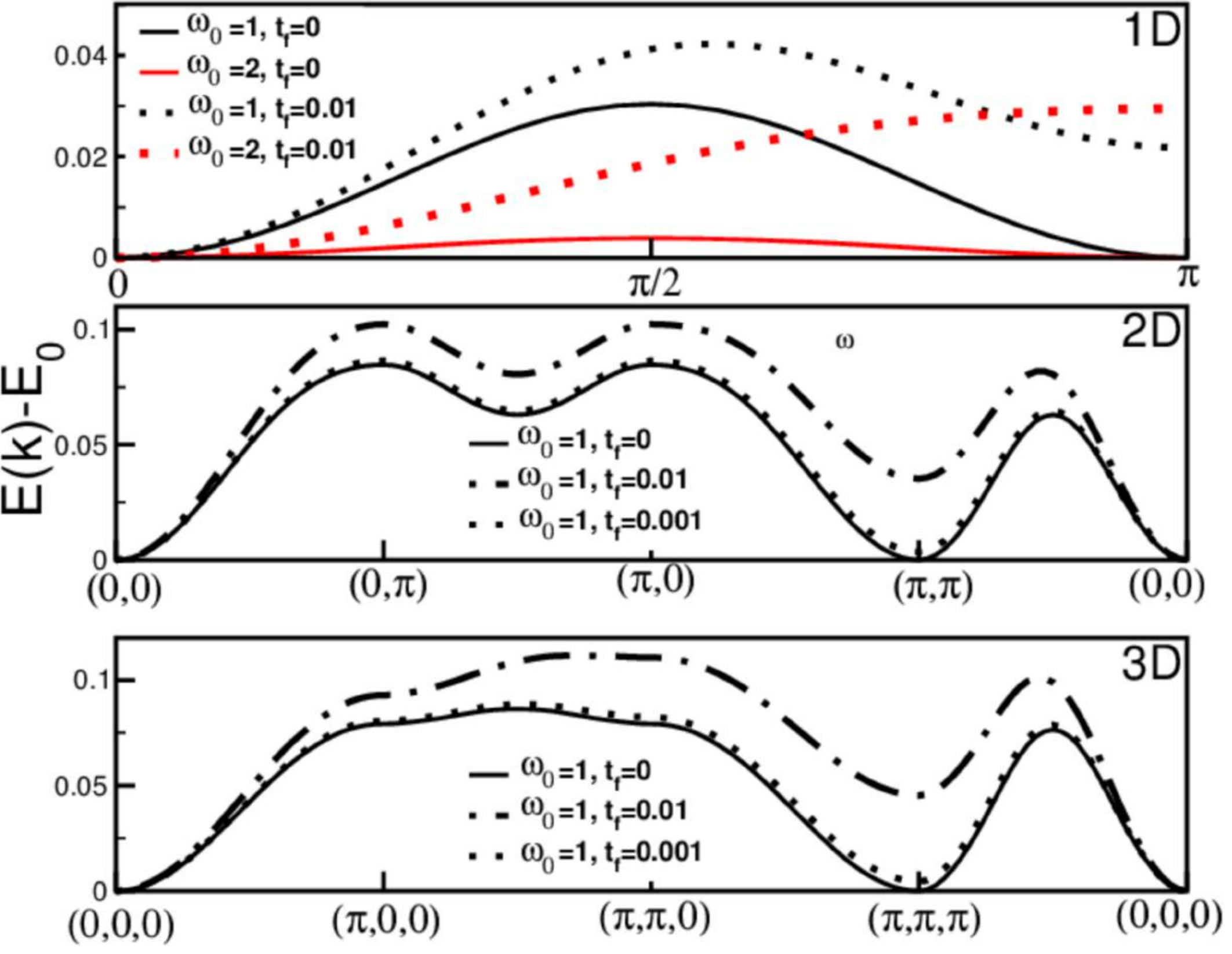}
\caption{(Color online)
Polaronic band dispersion ($E({\bf k})-E_0$) of the 1D, 2D, and 3D Edwards model in the small-$t_f$ large-$\omega_0$ regime.} 
\label{f3}
\end{figure}

\begin{figure}[t]
\vskip -0.0cm
\includegraphics[scale=0.39]{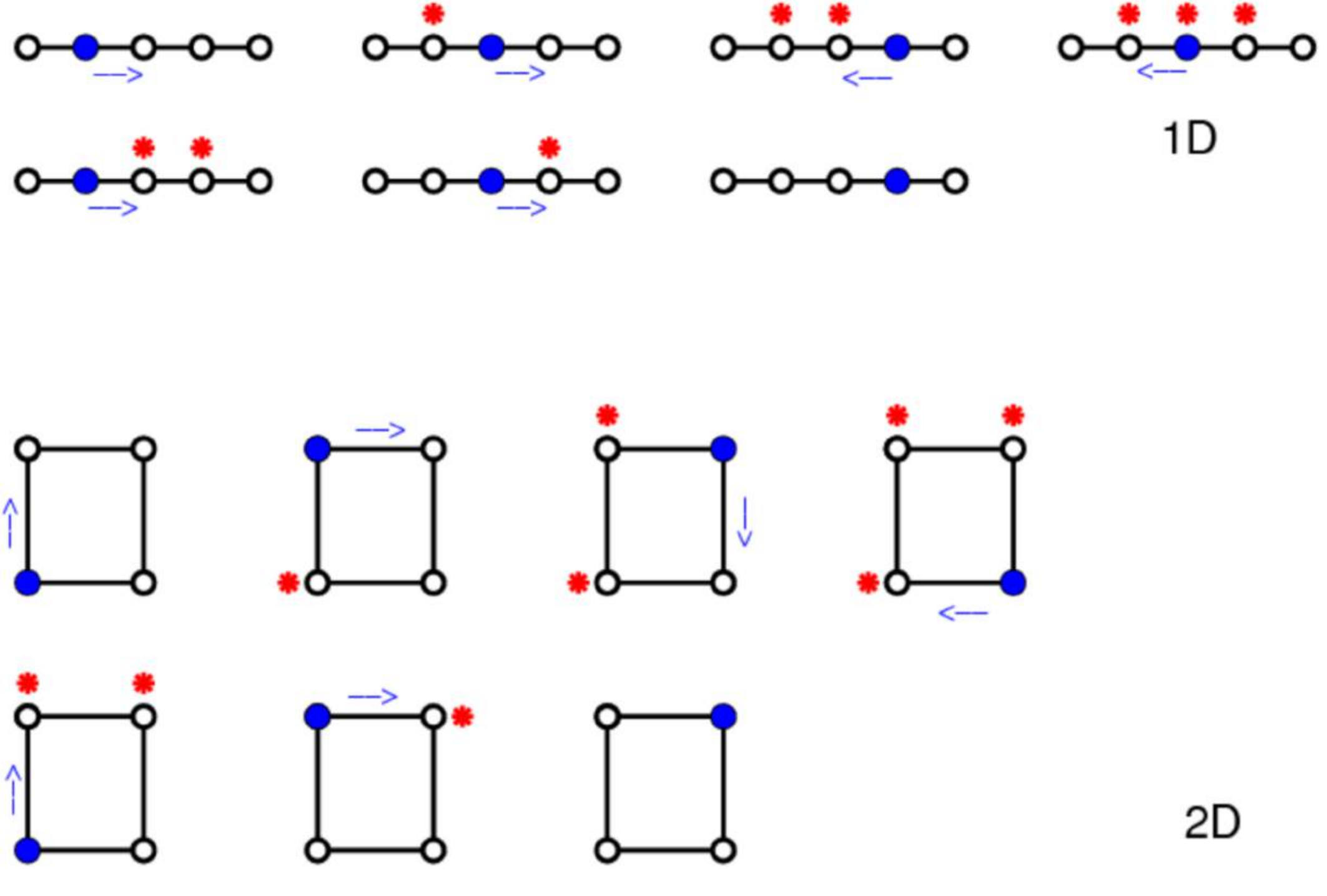}
 \hspace*{3.3cm}{\includegraphics[scale=0.50]{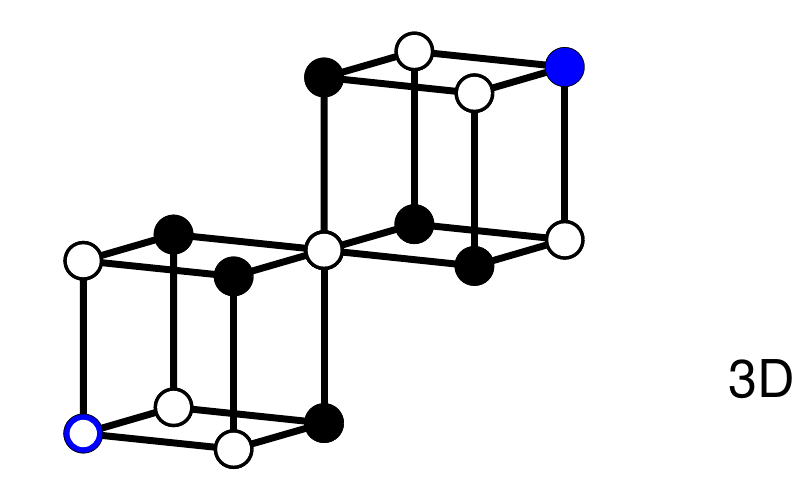}}
\caption{ (Color online)\label{fCL}
Sketch of the lowest order vacuum-restoring processes in the Edwards model. The top panel  illustrates the three-boson, three-site sequence of process that gives rise to an effective second nearest-neighbor fermion hopping in 1D~\cite{AEF07}.  The site occupied by a fermion is blue  and the arrow indicates the direction of the next $t_b$-hopping. Any fermion hopping either creates a boson (drawn as a red asterisk) at the site the particle leaves or absorbs a boson from the site it enters. While only collinear hops are allowed in 1D, collinear, noncollinear round the corner~\cite{BF10}, and closed loop  processes are allowed in 2D (middle panel). Note that, in 3D, any hopping process to the next nearest-neighbor body-diagonal site contains an odd number of hops and therefore is not vacuum restoring (both points belong to disconnected vacuum states). Vacuum-restoring hopping processes 
are only possible to the second nearest-neighbor body diagonal site, see  bottom panel.  Again these processes  are composed of 1D collinear, noncollinear, 
and/or 2D closed loop hops; they are of much higher ($18^{th}$) order in $t_b$, however.}
\end{figure}

\begin{figure}[t]
\hspace{-2.00cm}\includegraphics[scale=0.45]{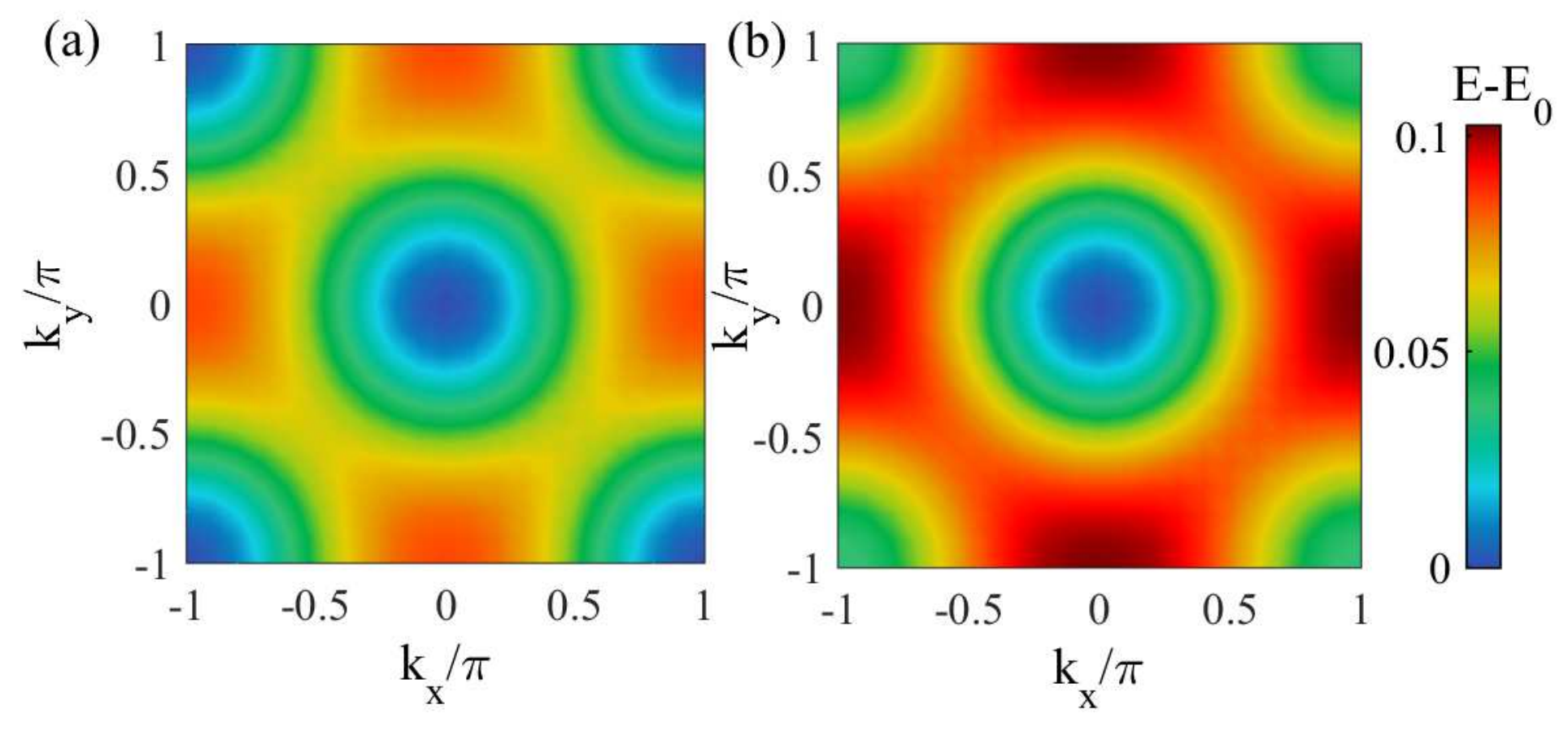}\hspace{-1.5cm}
\caption{ (Color online)
Ground-state energy shift, $E(k_x,k_y)$-$E_0$, as a function of $(k_x,k_y)$ for the 2D Edwards model with $\omega_0$=$1$ and $t_f$=$0$ 
 [note the band folding along the $(1,1)$-direction in reciprocal space]  (left), $t_f$=$0.1$ (right).} 
\vskip 0.5cm
\label{f4}
\end{figure}

\begin{figure}[t]
\vskip -0.0cm
\includegraphics[scale=0.5]{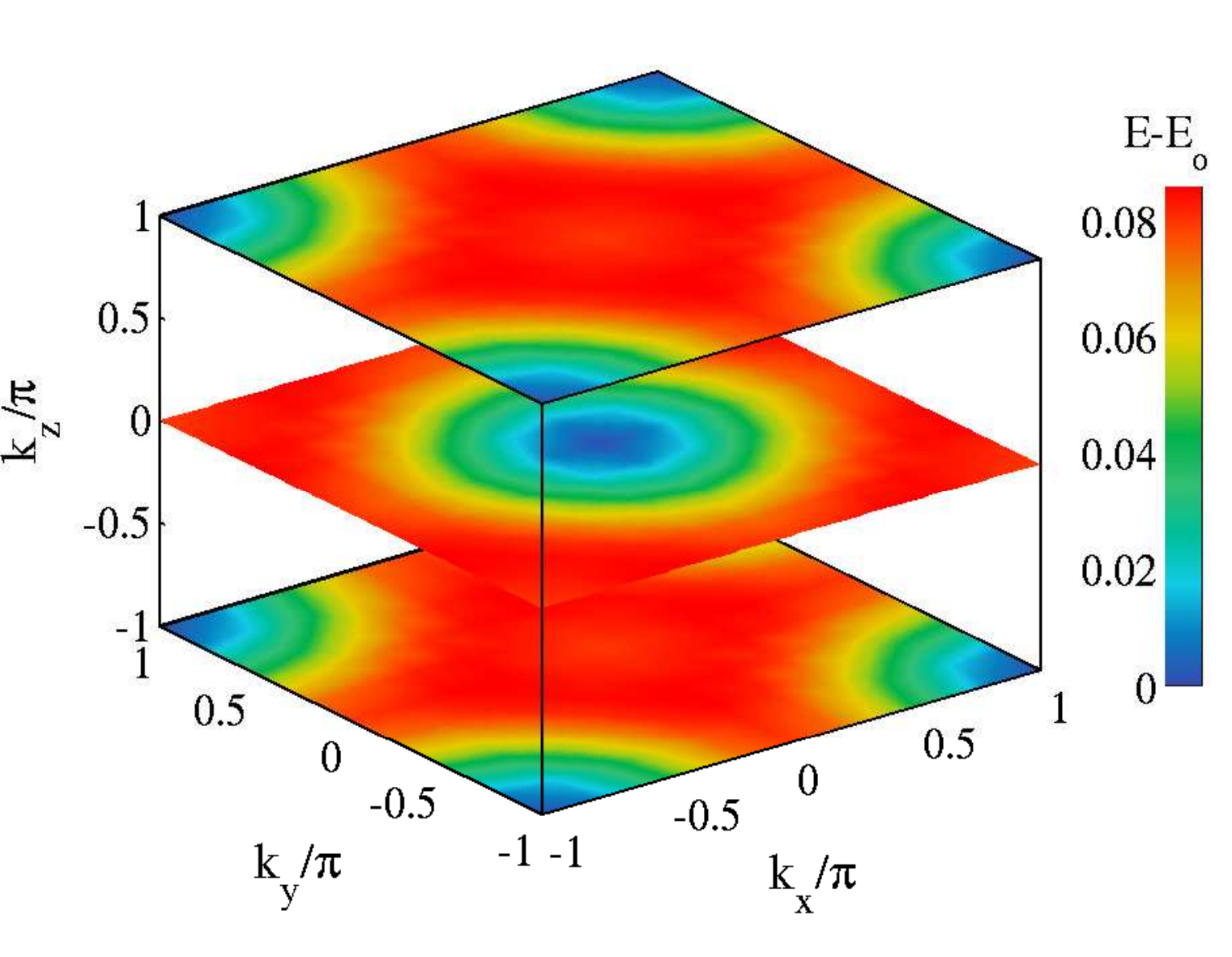}
\caption{ (Color online)
Contour plot of $E(k_x,k_y,k_z)$-$E_0$ as a function of $(k_x,k_y)$ at $k_z$=$0, \pm \pi$ for the 3D Edwards model. Again, $t_f$=$0$ and $\omega$=$2$ [note the band folding along the $(1,1,1)$-direction in reciprocal space].}
\label{f5}
\end{figure}

We first explore the quasiparticle energies $E({\bf k})$ in the Edwards model. Figure~\ref{f3} gives the polaron band dispersion in the regime where strong correlations in the background hinder the particle motion. Such a situation is realized at large values of $\omega_0$, where the bosonic excitations that are inherently connected to particle hopping are costly in terms of energy, and at small $t_f$, i.e., at small $\lambda$, when the ability of the background to relax is low. As a result the (coherent) bandwidth, defined  by the difference between the maximum and minimum  of  $E({\bf k})$ in the first Brillouin zone, is strongly reduced compared to the free particle's one. Clearly a ``true'' polaron band $E({\bf k})$ becomes only apparent if its bandwidth is smaller than the distance to the polaron-plus-one-boson continuum. In other words, the (lowest) quasiparticle band is well-separated from the incoherent part of the spectrum (or higher quasiparticle bands). 
This obviously is achieved in the parameter regime used for Fig.~\ref{f3}. We furthermore see that the polaron's bandwidth becomes larger as the dimensionality of the system increases from 2D to 3D. This is not difficult to understand because a string of bosonic excitations tends to bind the particle to the place where it starts its excursion. In higher dimensions, there are more ways to unwind such a string. Interestingly, coherent motion is nevertheless possible in 1D, and even for $t_f$=$0$, because there exists a six-step vacuum-restoring process~\cite{AEF07} which is a 1D representative of the 2D ``Trugman path''~\cite{Tr88} observed in a 2D N\'{e}el-ordered spin background, see Fig.~\ref{fCL}.  Since any hop of the particle changes the boson number by one, any vacuum-restoring process has to be in relationship to an even number of hopping events. It is worth noting that the strong correlations in the background medium give rise to a boson-modulated hopping that triggers a doubling of the unit cell (halving of the Brillouin zone). When $t_f=0$ (i.e., $\lambda=0$, and only vacuum-restoring hopping processes are allowed) the backfolding becomes perfect. This has been previously observed in 1D~\cite{AEF07} and 2D~\cite{BF10}. Figure~\ref{f3} demonstrates that $E({\bf k})$ is $(\pi, \pi, \pi)$-periodic also in 3D. The resulting dispersion reflects the developing many-body correlations  in the background medium. Since the coherent bandwidth (and the effective mass, see Sec.~III.C) is dominated by  the sequence of vacuum-restoring closed-loop hopping processes (which are two-dimensional in 3D as well, cf. Fig.~\ref{fCL}), the 2D and 3D bandwidths do not differ much.  The new periodicity of the Brillouin zone at $t_f$=$0$ is illustrated by the contourplots Fig.~\ref{f4} and Fig.~\ref{f5} for the 2D and 3D Edwards model, respectively. Of course, any finite $t_f$ will weaken the backfolding of the polaron band dispersion (see Fig.~\ref{f4}, right panel).

\subsection{Quasiparticle weight}
\begin{figure}[t]
\includegraphics[scale=0.4]{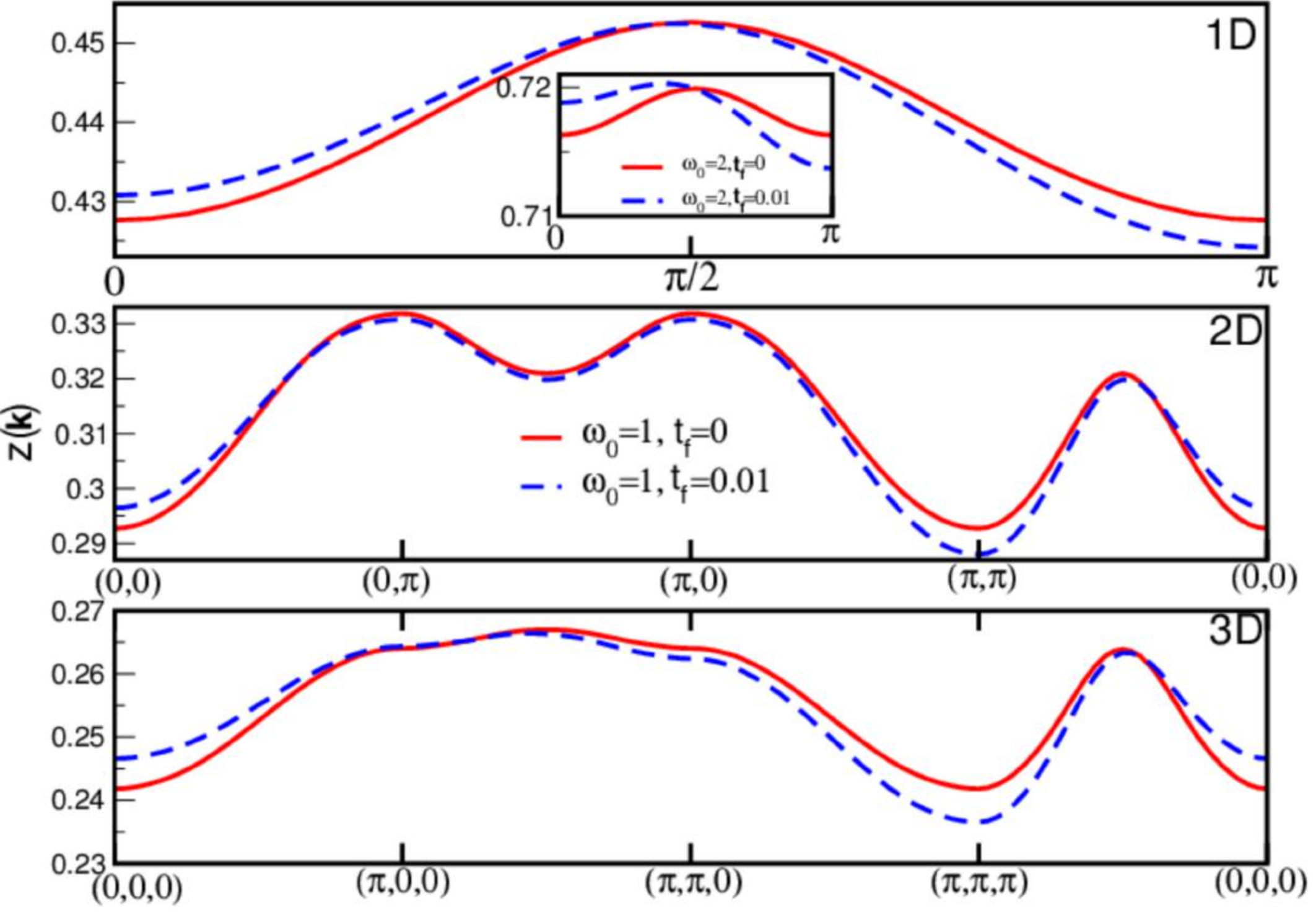}
\caption{(Color online)
Quasiparticle weight, $Z({\bf k})$, along the major directions of the Brillouin zone for
the 1D (top), 2D (middle), and 3D (bottom) Edwards model at $\omega_0$=$1$, and $t_f$=$0$ (red),
$t_f$=$0.01$ (blue). The inset gives $Z({k})$ at $\omega_0$=$2$ for the 1D case.} 
\label{f6} 
\end{figure}

Further information about the nature of the polaronic quasiparticle in the Edwards model can be obtained by computing the quasiparticle residue, 
\bea
Z({\bf k})=\vert\langle \psi^{(1)}_k \vert f_{k}^{\dag} \vert 0\rangle\vert^{2}\,,
\eea
which measures the overlap (squared) between the bare particle's band state $f_{k}^{\dag} \vert 0\rangle$ and the polaron ground-state wave function
$|\psi^{(1)}_k\rangle$~\cite{BTB99,BF10}. Figure~\ref{f6} gives $Z({\bf k})$ along lines of high symmetry in the Brillouin zone. First, we note that $Z({\bf k})$ is significantly 
reduced compared to free particle value (one). That is the Edwards polaron is heavily dressed by a cloud of bosonic excitations.
Even so, it is much less renormalized than the Holstein polaron~\cite{WRF96,WF98a,FT07}. Obviously, in the strong correlation regime, the Edwards polaron 
rather behaves as a spin polaron. Second, while $Z({\bf k})$ has a similar profile as $E({\bf k})$ throughout 
the Brillouin zone (cf. Fig.~\ref{f3}), it changes very little in real terms. This has been already observed for the 2D case within the momentum average approximation~\cite{BF10}, and retains it validity, as the exact data of Fig.~\ref{f6} indicate, in 1D and 3D as well. We note that at finite $t_f$, the quasiparticle weight is larger [smaller] at  (0,[0,0]) [$(\pi,[\pi,\pi])$] than the corresponding $t_f$=$0$-value. This is because the effective next-nearest-neighbor vacuum-restoring hopping process becomes less important if $t_f$$>$$0$.
 
\subsection{Effective mass}
Being able to calculate $E({\bf k})$ with high precision for continuously varying ${\bf k}$, we can compute the effective mass of the Edwards polaron for a $d$-dimensional hypercubic lattice, using the standard formula,   
\begin{equation}
 \frac{m}{m^*}=\frac{1}{d}\left[\sum_{i=1}^{d}\frac{\partial^2E(\vec{k})}{\partial {k_{i}}^2}\right]_{k_i=0}\;,
 \label{rem}
\end{equation}
where  $m$ denotes the ``reference'' mass, describing the a situation when both hopping channels are of equal importance (i.e., $t_f=t_b=1$). 

\begin{figure}[t]
\includegraphics[scale=0.4]{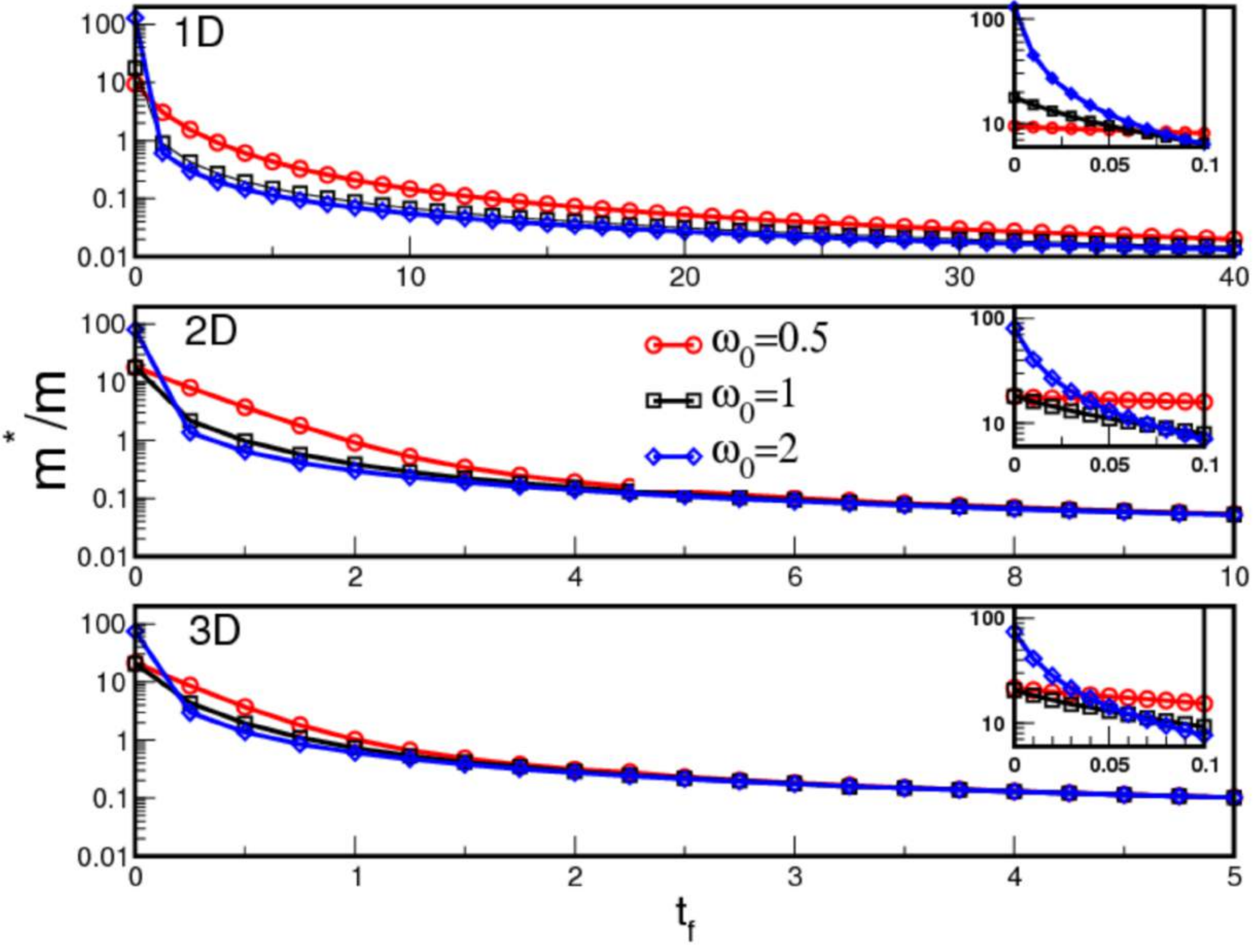}
\caption{ (Color online)
Effective mass $m^*/m$ in dependence on $t_f$ for the 1D, 2D, and 3D Edwards model (from top to bottom). Insets magnify the region of very small $t_f$.}
\label{f7}
\end{figure}

Figure~\ref{f7} displays the results obtained for the Edwards polaron's effective mass in 1D, 2D, and 3D. We first note that a finite $m^*$ results
even if the ```free particle'' has an infinite mass ($t_f$=$0$). The Edwards polaron transition is always continuous.  By contrast, in the SSH model, a sharp transition might appear when the coupling  depends not only on phonon momentum but also on the electron momentum~\cite{MDCBNPMS10}.
Considerable  differences are also observed compared to the Holstein model.   For example, the dimensionality affects the polaron crossover in a different manner (cf. the results given for the Holstein model in Ref.~\onlinecite{KTB02}). While the crossover becomes more defined  in higher dimensions for the Holstein case, the opposite tendency is observed for the Edwards polaron. Moreover, for the small Holstein polaron,  
the inverse effective mass obtained from Eq.~\eqref{rem} differs from $Z({\bf k=0})$ by less than 1\% ~\cite{BTB99}. 
As can be seen by comparing Figs.~\ref{f6} and~\ref{f7}, this difference is much larger (up to a factor of 100) for the Edwards polaron when $t_f\to 0$ in the strongly correlated regime.  In the case of boson-assisted transport, the dynamical generation of the effective mass is dominated by contributions from closed loops, which are comparably important in 2D and 3D (we already discussed in Sec. III~A that, in 3D, the lowest-order vacuum-restoring processes are basically the same as in 2D.)

Two more comments are in order here. 
First, in the ``diffusive'' or ``fluctuation-dominated'' transport regimes~\cite{AEF07} of small $\omega_0$, the mass enhancement is considerably  smaller. In this regimes, the quasiparticle band picture may even break down for the Edwards model (mainly because  $E({\bf k})$ is no longer separated from the polaron-plus-one-boson continuum). Second, as $t_f$ considerably exceeds $t_b$, we enter the quasi-free transport regime. Of course, for $t_f \to \infty$, $m^*$ (measured in units of the reference mass $m$) tends to zero. 

\begin{figure}[b]
\includegraphics[scale=0.4]{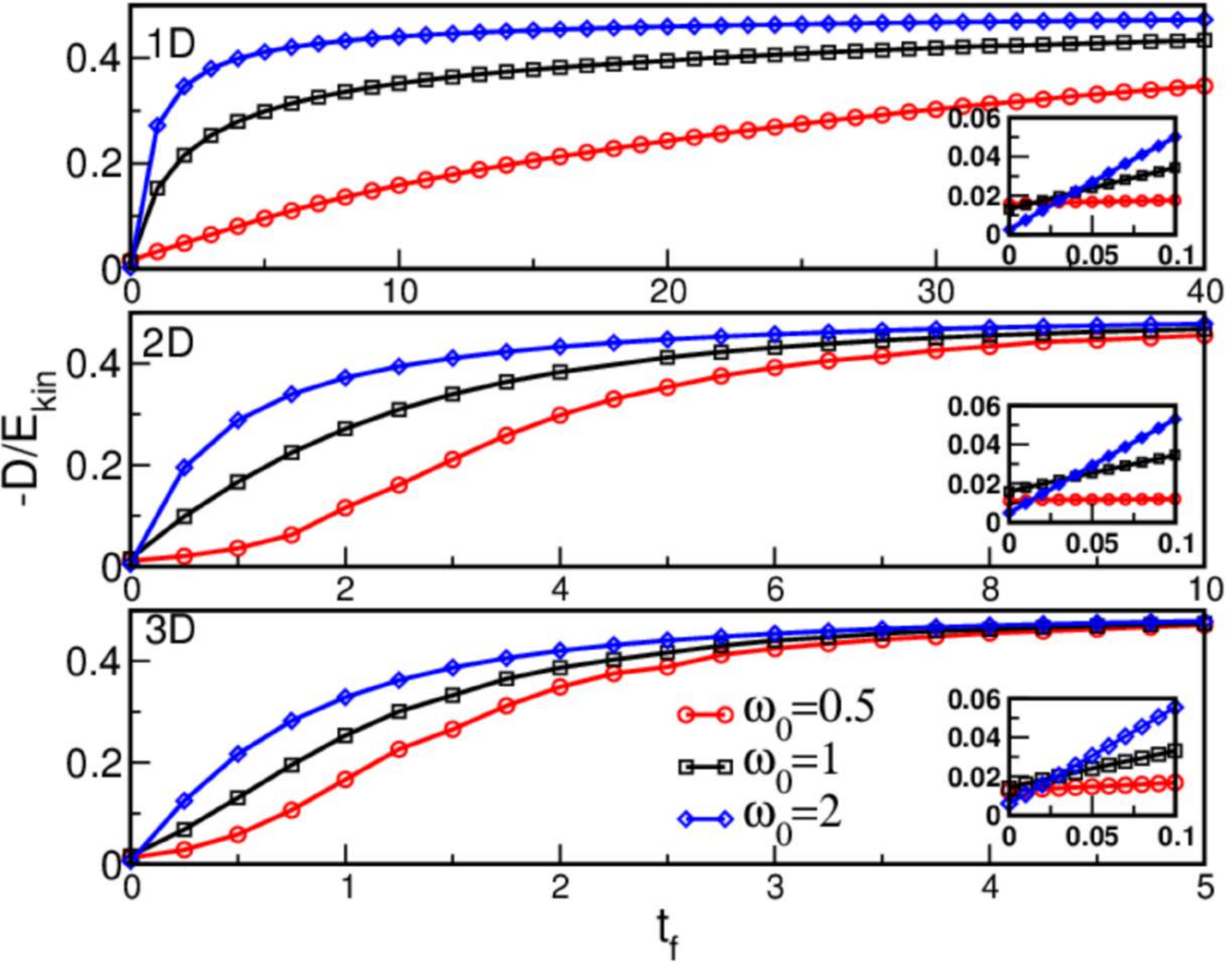}
\caption{(Color online)
Drude weight $D$ scaled to the kinetic
energy $E_{kin}$ as a function of $t_f$ for the 1D, 2D, and 3D Edwards model. Insets magnify the small-$t_f$ regime.}
\label{f8}
\end{figure}

\subsection{Drude weight}
In situations where electrical transport differs entirely from free particle motion, the Drude weight is {typically used to characterize transport}~\cite{Ko64,AEF10}. The Drude weight $D$ serves as a measure of coherent, free-particle like transport, and fulfills the $f$-sum rule. We have $-D/E_{kin}$=$1/2$ for a free particle, where $E_{kin}$ is the kinetic energy. By contrast,  $-D/E_{kin}\ll 1/2$ for diffusive transport. For our fermion-boson system, the Drude weight can be obtained by adding the same phase factor on the hopping matrix elements along the spatial directions of the hypercubic lattice  ($t_{f} \rightarrow t_{f} e^{i \phi}$, $t_{b} \rightarrow t_{b} e^{i \phi}$, which breaks time-reversal symmetry), and then exploit the relation~\cite{Ko64,PCT16}:
\begin{equation}
D=\left. \frac{\partial^2E_0(\phi)}{\partial \phi^2} \right|_{\phi=0}
\end{equation}
(in units of $\pi e^2$), where $E_0(\phi)$ is the ground-state energy in the presence 
of a non-vanishing phase $\phi$.

Figure~\ref{f8} shows the dependence of $-D/E_{kin}$ on $t_f$ at different values of $\omega_0$.
The 1D results are in excellent agreement with those of Ref.~\onlinecite{AEF07}. Here, the data for $\omega_0$=$2$ indicate that transport is quasi-free with $-D/E_{kin}\lesssim 1/2$ in a wide range of $t_f$. For $\omega_0$=$2$ and $t_f$=$0$, $D$ increases by about a factor of two (three)  in going from 1D to 2D (3D), which is basically due to the increasing coordination numbers of the corresponding hypercubic lattices. When $\omega_0$ decreases, the particle will be strongly scattered by background fluctuations, and $-D/E_{kin}$ tends to its asymptotic value 1/2 as $t_f\to \infty$ much slower. This characterizes the incoherent regime. On the other hand, for very small $t_f$, boson-assisted hopping is the dominant transport channel. Here $D$ increases with decreasing $\omega_0$ (see insets). Interestingly, for $t_f$=$0$, it can be shown analytically~\cite{AEF10}, that $D$ remains finite as $\omega_0\to 0$. 
These overall trends persist in 2D and 3D. However, there are subtle distinctions relative to the 1D case, for instance,  in the regime of small $\omega_0$: while $-D/E_{kin}$  stays almost constant for $t_f\ll 1$ when going from 1D to 3D, in higher dimensions,  it significantly exceeds its value at 1D for larger $t_f$ (note the different scales of the abscissae in Fig.~\ref{f8}). That  means, opening more hopping channels, the system approaches much faster the free-electron value in the diffusive regime (e.g., $D$ increases by a factor of 7-8, when going from 1D to 3D at $\omega_0$=$0.5$, $t_f$=$2$).

\begin{figure}[b]
\includegraphics[scale=0.4]{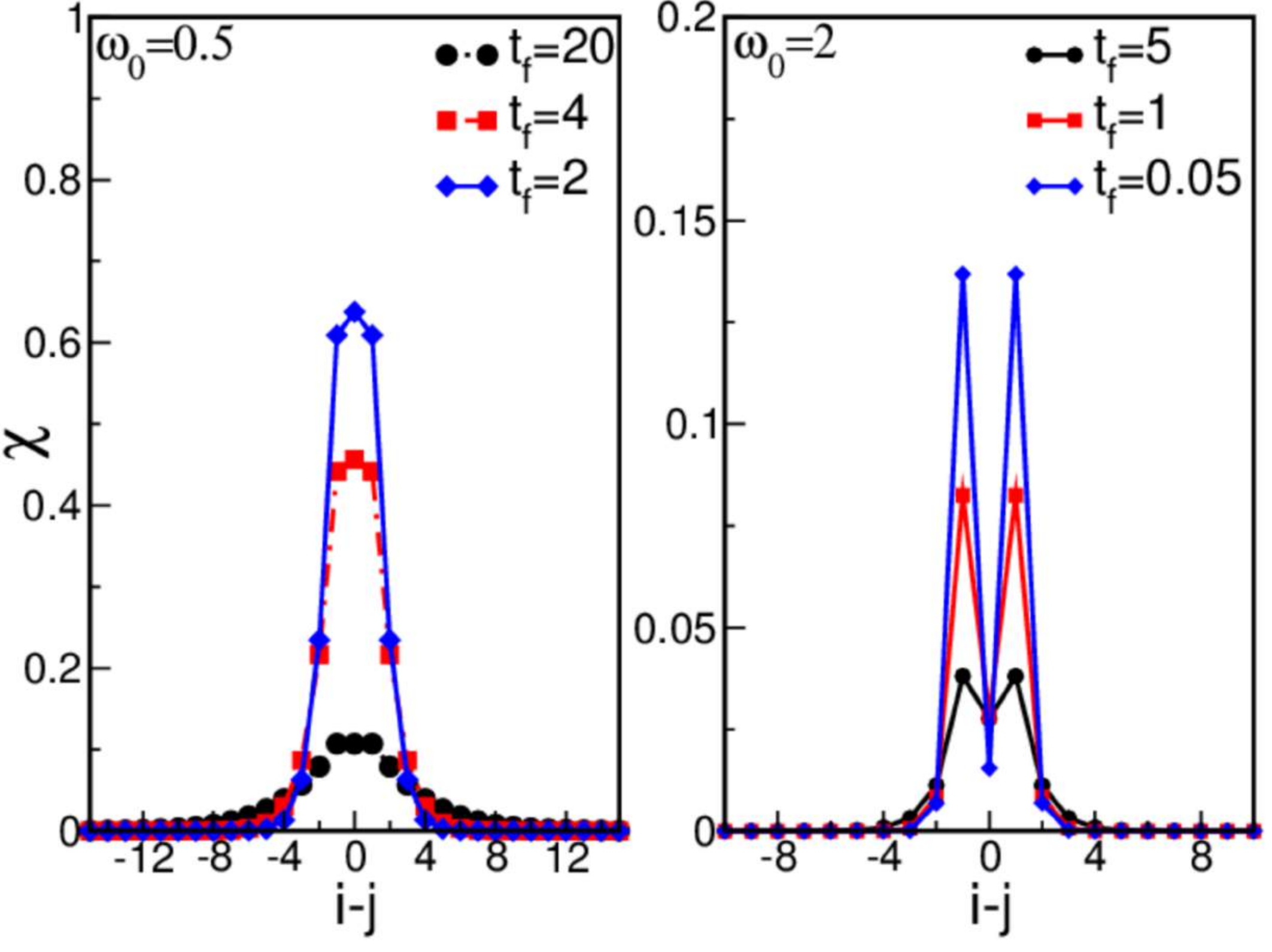} 
\caption{ (Color online) Particle-boson density-density correlation function $\chi(i-j)$ for the 1D Edwards model.}
\label{f9}
\end{figure} 

\subsection{Particle-boson correlations}
\begin{figure}[t]
\includegraphics[scale=0.5]{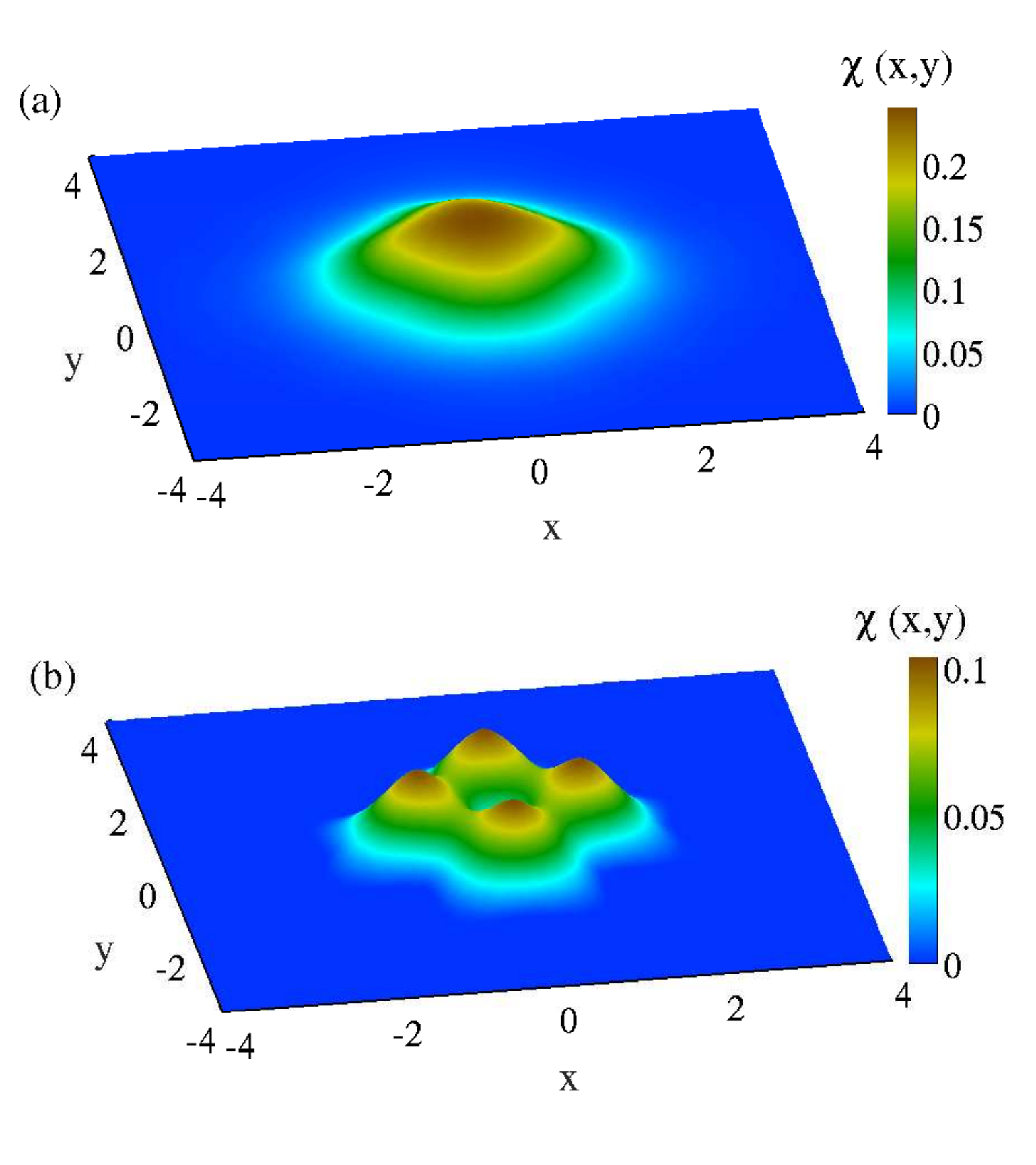}
\caption{\label{fy} (Color online) Particle-boson density-density correlation function $\chi(x,y)$ for the 2D Edwards model with $\omega_0$=$0.5$, $t_f$=$2$ (top), and $\omega_0$=$2$, $t_f$=$0$ (bottom).}
\label{f10}
\end{figure}

\begin{figure}[t]
\includegraphics[width=0.9\linewidth]{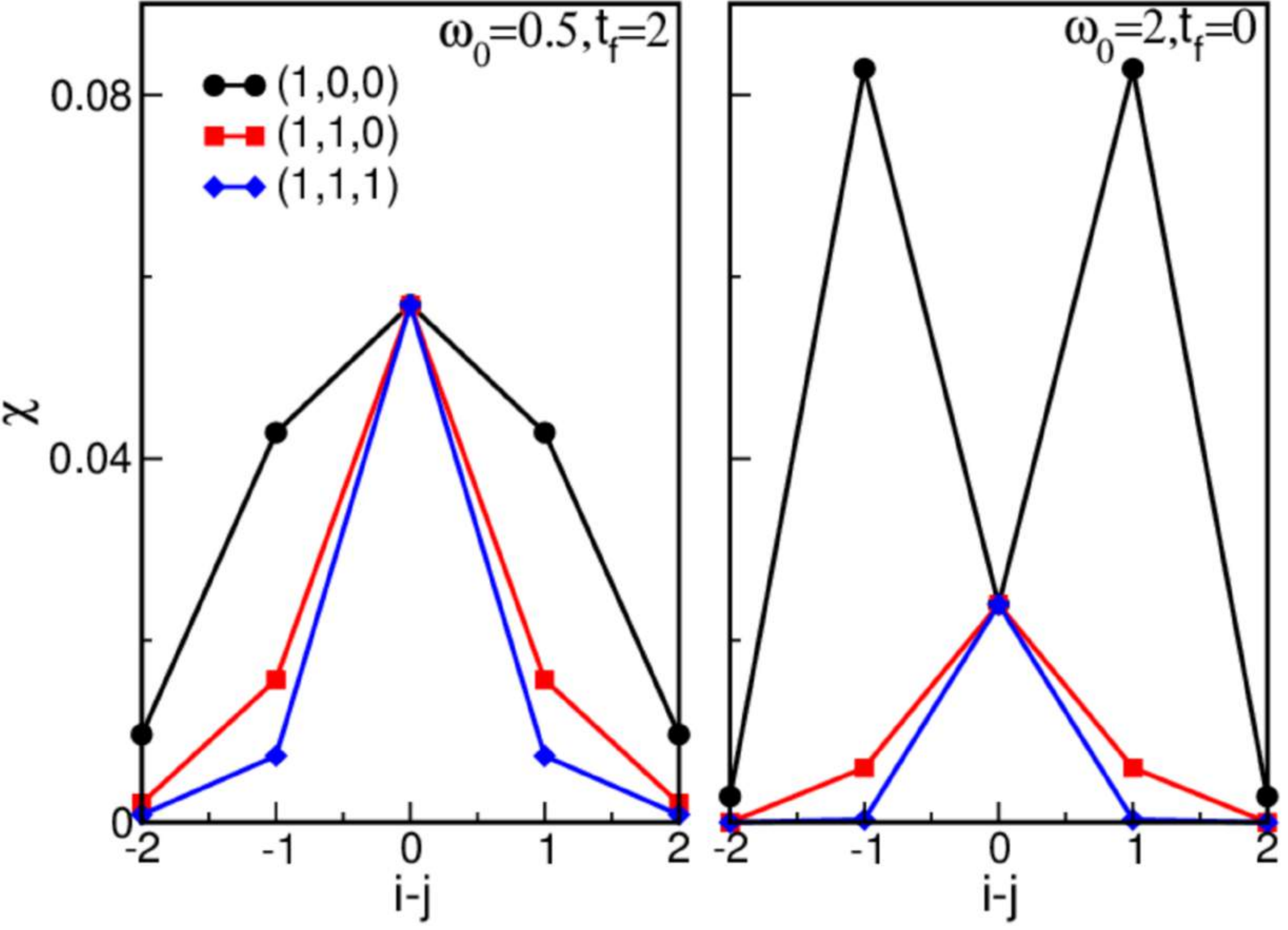}
\caption{\label{fy} (Color online) Particle-boson density-density correlation function for the 3D Edwards model 
with $\omega_0$=$0.5$, $t_f$=$2$ (left), and $t_f$=$0$ and $\omega_0$=$2$ (right). The distance from the particle-site is measured in lattice spacing along the (1,0,0) [black circles], (1,1,0) [red squares], and (1,1,1) [blue diamonds] 
direction.}
\label{f11}
\end{figure}

   The ground-state expectation value
\bea
\chi({\bf r})=\langle \psi_0 \vert f_i ^\dagger f_i^{} (b_{i+{\bf r}}^{\dagger} b_{i+{\bf r}}^{}) \vert \psi_0\rangle
\eea
captures the density-density correlation between the fermionic particle located at a certain site $i$ and the 
bosons in its proximity. Figures~\ref{f9}, \ref{f10}, and~\ref{f11} show  $\chi({\bf r})$  for  the one-, two-, and 
three-dimensional cases, respectively. In the incoherent, diffusive transport regime (i.e., at rather small $\omega_0$, $t_f > 1$), the bosons form a cloud surrounding the fermion. Here, the maximum of $\chi$ coincides with the position of the fermionic particle and the bosons are only weakly correlated. In total, many bosons are excited at the fermion site and in its neighborhood. To a certain extent, this resembles the situation for a large Holstein polaron. By contrast, in the boson-assisted transport regime, realized at large $\omega_0$ and very small or zero $t_f$, the particle-boson correlations are large at the nearest-neighbor sites. A boson existing on a site next to the particle triggers transport because, according to the second term in $H_b$, the particle can hop to this site and will thereby lower the total energy of the system by annihilating the bosonic excitation in the background. The same mechanism will strengthen hopping processes along the coordinate directions in higher dimensions too, whereupon, in 3D, transport along the body diagonal is not supported. This reveals once  more the importance of closed loops for the dynamical generation of the effective mass in the strongly correlated regime (cf. the results of Ref.~\onlinecite{BF10} for the 2D case). We would like to point out that the  nearest-neighbor particle-boson correlations are even more pronounced in 3D (and 2D) than in 1D (cf., the discussion of Fig.~\ref{f8} in Sec.~III~B).

\section{Conclusions}
To summarize, we have investigated the formation of polarons in the Edwards fermion-boson model, placing special emphasis on transport and dimensionality effects. The Edwards model features two transport channels, a coherent and an incoherent one. Exploiting unbiased (variational) diagonalization techniques, we presented numerically exact results for the Edwards model, including correlation functions and quantities that characterize transport, in spatial dimensions one through three. 

It turned out that an Edwards polaron mainly develops when the background is stiff (highly correlated).
Then coherent particle transport takes place on a strongly reduced energy scale. Entirely different from the Holstein- and SSH-type models, where the bosons are phonons and only (small) lattice polarons, comprising many phonons, will be formed (in D$>$1)~\cite{KTB02,FT07}, the Edwards polaron is a few-boson state in the regime of boson-assisted transport~\cite{AEF07} 
when vacuum-restoring processes play a dominant role. In that case, the Edwards polaron is confined to a few lattice
sites with pronounced nearest-neighbor particle-boson correlations. Edwards polaron formation requires a sizable  mass enhancement, just as
 in the case of Holstein- or SSH-polarons. Likewise, the Edwards polaron transition is always continuous, i.e., a crossover, triggered---in a self-induced way---by the strength of the background correlations. Interestingly, the inverse effective mass of the Edwards polaron substantially differs from the quasiparticle weight which, of course, is reduced from one, but rather moderate if compared to the Holstein polaron. For the dynamical generation of the Edwards polaron's effective mass, closed loops are important in all spatial dimensions. In the opposite limit, when the background heavily fluctuates, the particle will be strongly scattered by the bosonic fluctuations. This might enable transport when the ``free'' hopping channel ($\propto t_f$) is absent, but at the same time limits transport. In either case the Drude weight is finite, even if the energy of the background excitations ($\propto \omega_0$) tends to zero. We note that the limit $\omega\to0$ thoroughly differs from the adiabatic limit of the Holstein model~\cite{AFT10} (for the SSH model the polaron crossover is unaffected by the adiabaticity ratio~\cite{CSG97}). If, at small values of $\omega_0$, the ``free'' hopping channel is well-developed, the Drude weight (scaled to the kinetic energy) approaches its free-particle limiting value more readily in higher dimensions.   
Here, the boson cloud around the particle is spread but weakly correlated.   Obviously, the Edwards model captures
very different transport regimes, and the dimensionality noticeably affects the properties of the system  

Since the charge carriers in a rich variety of materials with strong electronic correlation, including 1D MX chains, 2D high-$T_c$ cuprates, and 3D colossal magnetoresistive manganates feature polaronic properties, our results contribute, at least qualitatively, to a better understanding of lattice, spin or orbital polaron formation in these materials, where particles move through an ordered insulator.

\acknowledgements
M.C. and H.F. would like to thank A. Alvermann for useful discussions.
The authors appreciate access to the computing facilities of the DST-FIST (phase-II) project installed in the Department of Physics, IIT Kharagpur, India. 
M.C. would like to acknowledge
funding from the NRF South Korea (No. 2009- 0079947), the POSTECH Physics BK21 fund, as well the computational facility at Department of Solid State Physics, 
Indian Association for the Cultivation of Science, Kolkata, India. B.I.M acknowledges the support from
National Research Foundation of Korea (Grants No.2015R1A2A1A15053564).
Work in Greifswald was supported by the Deutsche Forschungsgemeinschaft through SFB 652, project B5.

\bibliography{ref} 
\bibliographystyle{apsrev}
\end{document}